# Diagnosing Heterogeneous Dynamics in Single Molecule/Particle Trajectories with Multiscale Wavelets

*Kejia Chen,[†] Bo Wang,[§] Juan Guan,[§] and Steve Granick [†,§,‡,^,*]*

[†]Departments of Chemical and Biomolecular Engineering,

[§]Materials Science and Engineering, [‡]Chemistry, and [^]Physics, University of Illinois,

Urbana, IL 61801

*To whom correspondence should be addressed. E-mail: sgranick@uiuc.edu

**Abstract**

We describe a simple automated method to extract and quantify transient heterogeneous dynamical changes from large datasets generated in single molecule/particle tracking experiments. Based on wavelet transform, the method transforms raw data to locally match dynamics of interest. This is accomplished using statistically adaptive universal thresholding, whose advantage is to avoid a single arbitrary threshold that might conceal individual variability across populations. How to implement this multiscale method is described, focusing on local confined diffusion separated by transient transport periods or hopping events, with 3 specific examples: in cell biology, biotechnology, and glassy colloid dynamics. This computationally-efficient method can run routinely on hundreds of millions of data points analyzed within an hour on a desktop personal computer.

The experimental study of dynamics has been deeply transformed during the past generation by new technologies that acquire digital images in vast quantities, allowing one to record motion of objects of interest, one-by-one in real space and time.[1-6] When datasets of this kind are analyzed, the capacity to track individual objects over a long time allows not only quantification of individual variations within populations but also complex temporal fluctuations of individual moving elements. The valuable information offered by huge datasets goes beyond what can be obtained from the classic ensemble-averaged approach, and has often provided unexpected mechanistic insights. This approach of "deep" statistical imaging has already led to significant progress in a variety of fields, from physical sciences such as diffusion[5-10] and other dynamics in condensed matter,[4,11,12] to biological sciences such as ecology[13] and cell biology.[14-16] Much important work revolves around improving experimental techniques to collect the data.[17-18]

Here we ask a different question: how to analyze such data for embedded information? For many problems, but not all, it is reasonable to assume random fluctuations with some probabilistic distribution around an average value. However, dynamics in the physical and biological worlds are often heterogeneous. When the statistical character of the process changes intermittently with time due to stochastic switching between coexisting and often competing microscopic processes,[1-16] averaging over these distinct processes may give misleading results.

Progress is impeded by the paucity of methods to identify these distinct processes, and to quantify them, especially in the presence of noise in the data. An ideal method would be automated to handle large datasets, involve no judgment on the part of the analyst, and resolve rapid dynamic changes. In practice, to differentiate different modes of motion, one selects a metric of interest. Some of the metrics commonly used with aggregates of data include the scaling of mean square displacement (MSD),[19-21] correlation functions,[22,23] diffusion coefficient,[24] and other trajectory characteristics.[25-27] While each of these performs well for certain particular systems, these families of methods require prior information or assumptions about the character of the motion.

A delicate matter is to select the appropriate time window over which to seek dynamic changes of the observables. Too short a time can exaggerate noise but too long sacrifices temporal resolution and increases the chances of undesired mixing of distinct processes that switch more rapidly. Further, one needs to accumulate reliable statistics, but rare and transient events are too-readily averaged out, especially as reliability typically requires at least hundreds of data points. Also problematical is to select a criterion by which to distinguish random fluctuations from real changes in dynamics; there is no general way to avoid judgment in selecting these thresholds, and judgment risks being arbitrary, too subjective, and too demanding of different definitions depending on the case at hand. Taking a different approach, probability inference can be used to choose between models, maximizing the likelihood that a certain model fits the data.[21,24,28] But this presents its own complexity, especially because it requires prior assumptions in associating the data to particular models.

This paper presents wavelet transforms and universal thresholding as useful techniques. Developed in the field of digital signal processing,[29-32] here we show how to apply wavelet analysis to detect dynamic changes hidden in time-resolved positional trajectories of physical and biological systems. The advantage of this method is that it analyzes data on multiple scales simultaneously by decomposing time series into a full set of time scales while preserving information in time and frequency domains simultaneously. The following discussion presents first the method, then demonstrates its usefulness in three dynamical physical systems involving soft and biological matter.

**Methodology**

The qualitative principle of wavelet analysis is simple:[29-31] moving along a time series of observables, transient changes in dynamics result in wavelet coefficients with large absolute values, as wavelet transform is known to be very effective at detecting discontinuity. Fig. 1a shows schematically the main idea: a time series of raw data is expanded to time-resolved wavelet coefficients on different scales (or frequency), by convolution over local times with the wavelet basis function. Background "noise", which represents in part random fluctuations, in part the mixing of different processes in the system, is measured on small scales (high frequency), and projected statistically to larger scales (low frequency) generating a "universal threshold" that naturally adapts to the noise amplitude. This threshold allows one to discriminate dynamics of interest, "signal" that exceeds this threshold, on larger scales.

One inspects a time series of an observable (Fig. 1b). There results a spectrum of wavelet coefficients against time and scales (Fig. 1c). Over small scales, the wavelet coefficients are dominated by featureless random fluctuations, whereas at large scales, the coefficients of given time points are heavily distorted by dynamics extending for long times around it. Importantly, at the intermediate scales, transient changes above background correspond to distinct bands in wavelet coefficients that can be resolved with confidence. Detecting the convergence to these local maxima of wavelet coefficients on these scales localizes dynamic heterogeneity, the information we seek. As explained below, this local detection has a time resolution better than the exact scale on which this analysis is carried out. This statistically rigorous multi-scale method overcomes the current difficulties in analyzing heterogeneous dynamic data as we have introduced.

To implement the method there are 4 steps: (a) choose the wavelet basis function; (b) perform the wavelet transform; (c) determine the scale and threshold; (d) assign the physical processes.

**Wavelet selection**. The wavelet transform calculates the local integral values of time series over various scales with weighting defined by the wavelet function used. Different scales generate a times series of wavelet transform coefficients corresponding to different time scales. Specifically, the wavelet transform[29-31] of a time series $s(t)$ on scale $a$ is

$$C(t_0,a) \equiv \frac{1}{a}\int_{\mathbb{R}} s(t)\psi\left(\frac{t-t_0}{a}\right)dt \qquad (1)$$

where the wavelet function $\psi$ has width $a$, centered at time $t_0$. To satisfy Lipschitz continuity, local maxima of $C(t_0,a)$ correspond to singularities in $s(t)$. Therefore, to detect dynamical changes, one searches for regions converging to local maxima in $C(t_0,a)$ along $t_0$.

The appropriate wavelet depends on the nature of the dynamic heterogeneity for which one searches. For example, if the background contains fast fluctuations, then one selects $\psi$ with one vanishing moment (such as Haar wavelet), such that local maxima in $C(t_0,a)$ correspond to discontinuity above this constant fast fluctuating component. As another example, when the dynamics exhibits abrupt, fast back-and-forth jumps between several positions (*e.g.* harmonic oscillators in double potential wells), the detection should target maximum curvature in the trajectory. Then one selects $\psi$ with two vanishing moments (such as Symlet and high order Daubechies wavelets) such that local maxima in $C(t_0,a)$ correspond to maximum curvature in trajectories.[29]

A third example is the one on which we focus in the three physical examples discussed in detail below: this is the case of local confined diffusion separated by transient transport periods or hopping events. The appropriate wavelet should have one vanishing moment. This, the "Haar wavelet", quantifies the displacement between the average position of $n$ points before and after position $i$ along a trajectory with equal weighting to the position of all the points around the center point:

$$C(i, 2n) = \frac{1}{n}\left(\sum_{j=i}^{i+n} x_j - \sum_{j=i}^{i-n} x_j\right). \qquad (2)$$

What this means physically is quantifying the drift of mean position over time. For Brownian motion, there is no drift of the mean position, which oscillates around zero. But for directional transport, drift is decidedly finite.

We emphasize that the choice of wavelet must be physically driven, and that different wavelet functions can be more appropriate depending on the problem at hand. While the mathematical steps of implanting this method are straightforward, as a premise one needs judicious judgment of what type of motion to target.

**Implementing the wavelet transform**. The wavelet is mathematically defined as local integrals according to Eq. (1), but in practice it is more efficient computationally to compute the coefficients by correlations between a short section of the time series data and the chosen wavelet, shifting and stretching the wavelet according to $t_0$ and scale $a$. Using MATLAB for convenience, we compute continuous wavelet coefficients at real, positive scales of trajectories projected onto x and y dimensions separately. An example is shown in Fig. 1. This shows the result of a trajectory transformed into time series of coefficients over various scales.

**Choosing the appropriate scale and threshold**. One must set a scale on which a threshold is used to decide what differences are large enough to matter. As a practical protocol, we find it convenient to select scale and threshold with the following iterated sequence: we typically make an initial guess of the scale, $\tilde{a}$, on which distinct bands of wavelet transform coefficients start to emerge (Fig. 1b). Then we use what is called "universal thresholding" on this scale.[30] For the chosen scale $\tilde{a}$, the threshold is projected from scale $a = 2$ using

$$\delta = \eta \sigma_2 \sqrt{2 \ln N} \qquad (3)$$

where $N$ is the number of data points in this trajectory excluding first and last $\frac{\tilde{a}}{2}$ data points, $\sigma_2$ is the estimate of the standard deviation of noise on scale 2, and $\eta$ is the projection factor. To detect persistent transport above Fickian diffusion, we use $\eta = \sqrt{\frac{\tilde{a}}{2}}$ because random noise arising from Fickian diffusion grows with $\sqrt{t}$. As coefficients on scale 2 are a mixture from multiple processes, the standard deviation of "noise" cannot be calculated directly from the data, so we estimate $\sigma_2$ using the median absolute deviation (MAD)[30]

$$\sigma_2 \equiv \frac{\text{median}_i\left(|C(i,2)|\right)}{0.6745} \qquad (4)$$

Finally, we refine the scale and threshold through iterated assignment and validation of training trajectories, if necessary. Criteria by which to decide whether refinement is needed are (a) whether the assignments are insensitive to small changes of scale and threshold, and (b) whether other metrics, such as MSD and correlations, are separated in ways that are anticipated based on physical characters of these processes.

Note that while this detection method is local, the estimate of noise level is global, as the entire trajectory is used to estimate the noise level. This gives reliable and fully automatic thresholding adaptive to each individual trajectory, with thresholds reflecting the heterogeneity between trajectories. Moreover, in implementing this method, we have noticed that a wide range of scales give equally robust separation, as will be illustrated below.

**Interpreting data to assignphysical process.** Naturally, the physical process of interest depends on the problem at hand. In the sections below, we present three examples of distinguishing between "signal" and "noise." In all three examples the "signal" contains heterogeneity that is characteristic of the system, and the noise describes the Brownian component that constantly fluctuates as background.

Assignment proceeds by combining the time periods during which wavelet coefficients on scale $\tilde{a}$ exceed the threshold $\delta$. For multidimensional trajectories, *e.g.* x, y, and z in Cartesian space, this is repeated for each dimension and the results are combined. When the trajectories are isotropic, a common threshold can be used for all dimensions. The confidence of assignment depends on the trajectory length, especially for very short trajectories, as then a reliable estimate of noise level becomes impossible. We typically discard trajectories shorter than 300 data points.

## Three Examples

To test the efficacy of this method and to illustrate the operation in practice, we illustrate this wavelet analysis with three examples: to distinguish active transport from passive diffusion of single-particle motion in cell biology, to identify pauses of single-molecule DNA trajectories in electrophoretic mobility, and to capture intermittency of single-particle glassy dynamics. Focusing on the first example, the latter two examples illustrate that this method is general.

**Active transport in living cells**

Intracellular transport of endosome "cargo" proceeds by a stochastic switching between passive diffusion and active transport along microtubules, dragged by motor proteins, kinesin and dynein.[33] This switching between two types of motions, active and passive, is known to happen on sub-second time scales; in function, it enhances cell reaction kinetics and maintains cellular functions.[34] It is of fundamental significance to resolve active processes uncontaminated by passive fluctuations, but to do so, methods are needed to discriminate between them.

The data are contained in a Ph.D. thesis.[35] Using fluorescence microscopy (see experimental details below), we obtained trajectories of EGF-containing endosomes in living HeLa cells and we implemented wavelet analysis to assign passive and active motion. From raw data of trajectories illustrated in Fig. 2a and Fig. 2b, two types of motions cannot be distinguished on short time scales. We computed the Haar wavelet coefficients at a scale of 32 frames (Fig. 2c), a scale on which banding of wavelet transform coefficients is clearly distinguishable. The threshold was set according to universal thresholding with results indicated in gray in Fig. 2c, the segments that exceeds the threshold were assigned as active. Combining results both on x, and y, the separation is overlaid on the original trajectory in Fig. 2a, with active portion highlighted in red. Fig. 2d shows that the assigned active segments were invariably super-diffusive while passive segments were invariably either diffusive or sub-diffusive, which is reasonable physically.

The major advantage of wavelet analysis, for this example, is considered to be that universal thresholding is adaptive to heterogeneities between trajectories. We found the imputed thresholds to vary by two orders of magnitude, depending on the rate of passive motion of that particular trajectory (Fig. 3a). Nonetheless, Fig. 3b shows that according to the trajectory, the imputed active fraction spanned a wide range that was independent of the threshold, suggesting that thresholding did not introduce an artificial bias. The arrows point out trajectories, which are <10% of all, for which after the initial assignment, fine-tuning of the threshold was necessary. They were the trajectories with a prohibitively large fraction of active transport to estimate the passive fluctuations using MAD (Eq. 4). They were identified during the refinement step where for each individual trajectory, the mean square displacement (MSD) of the assigned passive portion was calculated and fit as a power law in time, $\langle \Delta r^2 \rangle \propto t^\alpha$, where $\alpha < 1$ is expected for passive motion according to physical reasoning. For these trajectories, the fitted $\alpha$ initially exceeded 1.1, and the thresholds were decreased and the process repeated iteratively until the criterion $\alpha \leq 1.1$ was satisfied.

To test the accuracy of the assignment, we disrupted microtubules using nocodazole, which eliminates active transport, and then the trajectories were analyzed as before. Fewer than 0.1% of total steps were mistakenly assigned as active motion. To further test the reliability of the wavelet-based assignments, manual checks were performed. We inspected 10 representative trajectories, a total of 16,816 image frames, and manually assigned the active frames. This visual inspection found that false-positive of active steps amounted to only ~5% of the total active steps. Also, visual inspection suggested that ~20% of the active steps were missed by the wavelet analysis. Similar performance was confirmed on simulated trajectories (Methods Section). As visual inspection was subjective, we are unable to decide to which method more confidence should be given. Our main conclusion is two-fold. First, errors of the wavelet analysis tendedto err on the conservative side, tending to mistakenly assign active steps as passive motion. Detailed arguments concluded that these miss-assignments did not bias the data.[35] Second, this conservatism resulted in excellent discrimination of the active steps themselves.

The visual inspection suggested that miss-assignments occurred at the transition of active and passive motion. Other false negatives identified manually are more debatable. For example, sometimes the active motion circled around or reversed directions as shown in Fig. 4. Depending on the dynamics of interest, one may or may not want to identify these motions. The Haar wavelet will assign them as passive (highlighted by arrows in Fig. 4a) as the drift of mean position approaches zero at those points; while Daubechies 4 wavelet with 2 vanishing moments, which targets motions that introduce abrupt changes in local curvature, will assign them as active (Fig. 4b). So an appropriate wavelet should be selected based on the need.

Advantages of this analysis are considered to be mainly two. First, large datasets were processed automatically in a short time. Visual inspection would have been prohibitive; the visual inspection of 10 trajectories just mentioned consumed roughly 5 hr, whereas on a PC 3443 trajectories were analyzed in 10 minutes. Second, the wavelet analysis suggested significantly new conclusions about the physical process, identifying that EGF-containing endosomes in HeLa cells spend around 30% of the total time (879,427 out of 3,211,776 steps) in active transport, whereas the fraction found by existing methods reported in the literature[20,22,23,25] was less than 5% with more false positives (see Methods Section for details). Furthermore, the average duration of continuous active transport was assigned to be ~0.75 s using wavelet analysis but less than half of this (~0.3 s) using the other methods.[20,22,23,25] These differences are considered to be a significant advantage of the method.

**Intermittent mobility in DNA electrophoresis**

Recent single-molecule measurements from this laboratory show that when λ-DNA migrates through agarose gel under the action of an electric field, the time-dependent position of individual molecules proceeds in spurts with pauses in between.[36] This is another problem of how to separate "signal" (the spurts) from "noise" (the pauses), in the presence of uninteresting background noise. While to the eye it may be obvious that molecular mobility exhibits two states (Fig. 5a), to automatically separate these two is challenging, as the frame-to-frame displacements of center of mass shows no temporal pattern above random fluctuations (Fig. 5b).

To discriminate these two mobility states, a wavelet analysis was used on each single-molecule trajectory. On scales 8 and 32, the universal threshold separated the pauses and jumps nicely for λ-DNA in 1.5 wt% agarose gel under electric field 12 and 6 V/cm respectively (Fig. 5a and c). At each field strength, a range of scales give good separation (Fig. 5c). Too-small scales result in missing a large portion of spurts, too-large scales assign mistakenly many pauses as spurts. The envelope of usable scales (which still spans a broad range) decreased with field strength because, as the lifetime of pausing shortened as the force on the DNA molecules increased, shorter scale became better at most accurately assigning these faster transitions.

Analyzing ~1,000 trajectories we found that spurts comprised 30% (68,222/230,057 steps) and ~60% (61,436/102,574 steps) of the total time elapsed, under electric field 6 and 12 V/cm respectively. The speed during spurts much exceeded the mean speed (Fig. 5a). This significant contrast would not have been quantified otherwise, and provides firm numbers from which to examine competing electrophoresis theories.[37]

**Hopping dynamics in a colloidal supercooled liquid**

It is well established that dynamics in supercooled liquids is intermittent.[11,38,39] As illustrated in Fig. 6a, in both positional and angular space, trajectories are at first restricted to a narrow region of space, then hop suddenly to a new region, probably reflecting collective rearrangements of the neighbors.[39] Despite numerous previous studies, such hopping events are hard to identify automatically in large datasets. The difficulty is that, during hopping, displacements or directional persistence of the motion do not necessarily differ from those during caging (Fig. 6b), and the duration of the hopping is typically short, if not instantaneous. Therefore, these events are often characterized ambiguously as "fat tails" of total ensemble-averaged displacement distributions.[5,11,38]

However, these abrupt changes present pronounced peaks in wavelet transform coefficients (Fig. 6c). By the methods described above, they can be located precisely by a wavelet analysis. Detecting hopping in this way, one notices hopping simultaneously in position and rotation, indicating that rotational motion correlates closely with translational motion. This observation agrees with our previous conclusion drawn from ensemble-averaged correlation functions,[39] but from the wavelet analysis the evidence of rotational-translational coupling is made more direct.

As this transition is so sharp, and presumably instantaneous, this example poses a stringent test for the time resolution of our method. Fig. 6d shows that the time resolution improves when the scale becomes smaller and the threshold increases. However, the difference is small, suggesting that the time resolution is related to but is not limited by the scale on which the analysis was carried out. As a rule of thumb, a wide range of scales give similar time resolution, ~5 frames. Further, by changing the time between evaluations of the data, we showed that the time resolution is largely unaffected. These tests demonstrated the robustness of the method and the ability to identify truly transient dynamics.

With this method, it would be interesting to revisit the enormous amount of data that is available in the literature, [5,11,38] to directly measure the caging time, as well as its distributions, when approaching the glass transition. Further, using these hopping events as reference points, one could examine the dynamic paths and structural organizations around those events. Information of this kind could be difficult to obtain otherwise, as rare events of this kind can be buried deeply in conventional correlation functions that average over all time.

**Conclusion**

We have described a robust method to automatically detect dynamic heterogeneity in time series data that are collected routinely in many labs across various fields. We have applied this method to three separate examples: in cell biology, biotechnology, and soft matter physics, to illustrate and validate the method, and to demonstrate its broad usefulness. Since the analysis makes no assumptions about the physical nature of the dynamics, we emphasize that the method is general.

It is impossible to achieve perfect separation without full knowledge of the microscopic mechanism. One only can do so with a certain level of statistical confidence, depending on the method used. Apart from the wavelet method described in this paper, the existing methods fall into two categories. One class is based on fitting the data to presumed models. A second class is based on ensemble-averaged quantities which suffer from incompatibility between time resolution and statistical reliability. With 3 applications discussed above, and in another test of the method using simulated data whose statistical character was known precisely by direct input of the data, we have discussed how a wavelet approach can aid in going beyond these limitations.

### Methods

**Active transport.** Endosomes in HeLa cells were fluorescently labeled by incubating the cells with 0.15 µg/mL biotinylated EGF complexed to Alexa-555 streptavidin (Invitrogen) for 20 min. The endosomes were tracked under physiological conditions on a homebuilt microscope in a highly inclined illumination optical (HILO) geometry with the laser beam inclined and laminated as a thin optical sheet with thickness of ~1 µm into the cells.[40] To avoid focal plane drifting, the objective was simultaneously heated and immersion oil with ultralow fluorescence standardized at 37 °C (Cargille) was used. Fluorescence images were collected by a back-illuminated electron multiplying charge-coupled device (EMCCD) camera (Andor iXon DV-897 BV). Typically, each movie lasted 4,000 frames at a frame rate of 20 fps. The movies were converted into digital format and analyzed using single-particle tracking programs,[41] locating the center of each particle in each frame and stringing these positions together to form trajectories. The tracking uncertainty was < 5 nm.

**Active transport simulation.** To validate wavelet analysis on active transport, we simulated trajectories for which the true statistics are known. The following properties were fed into simulations: 1) the active transport had an exponential distribution of step size, the average being the experimentally-measured value, ~40 nm/step (50 ms per step); 2) the direction between adjacent active steps was allowed to vary by an angle selected at random from a uniform distribution bounded by π/50 to mimic the upper limit of the curvature observed in experimental trajectories (~ 1 μm); 3) fluctuations perpendicular to linear active transport were introduced as a Gaussian noise with width equal to the experimentally-measured value of 40 nm; 4) passive motion was simulated so as to generate sub-diffusive MSD curves similar to those we observed in cells whose microtubules had been disrupted by nocodazole; this was accomplished by positioning the steps randomly within an area defined by a 2D Gaussian spreading function centered at the average position of the previous 50 passive steps with width of 100 nm; 5) the transition between passive and active motion was assumed to be Poissonian with transition probabilities set to reproduce the observed average length of active runs, which was 20 steps (~1 s), the total active portion being ~20%.

**Implementation of existing methods that detect active transport.** To compare the performance, we implemented three existing methods in the literature which uses speed correlation,[22,23] asymmetry,[25] slope of MSD and standard deviation of angle correlation[20] as the characteristics to define active transport. These quantities were calculated for each point in the trajectory using a rolling window. The rolling window size was estimated as described.[20,22,23,25] The window sizeneeds to be long enough for statistical significance, but shorter than the duration of the active transport. Since the average duration of active motion is estimated to be around 1 s using the wavelet analysis, the window size was selected to be 11 in our analysis, which is about half of the average duration. The odd number was used to allow equal number of points before and after the point of interest in the rolling window. The $L_{max}$ for the asymmetry method was set at 71 to match the longest possible duration of active motion.

The thresholds for all three methods were determined from Brownian simulation. 100 Brownian trajectories of N=1000 frames with 20 fps were simulated for diffusion coefficient D =

0.001μm²s⁻¹. The trajectories were composed of steps with a uniform probability distribution for step direction and an exponential probability distribution for step length with a mean of $\sqrt{4D\Delta t}$. According to the literature,[20,22,23,25] the threshold was defined for each parameter so that 99% of the simulated trajectories were classified as passive. These thresholds were designed to include 1% false positives, but the real performance was worse according to our validation test. The thresholds for speed correlation, asymmetry, slope of MSD and standard deviation of angle correlation were 0.886, 1.25, 2±0.4 and 1.1 respectively. We verified that the thresholds were sensitive to neither N nor D. When comparing the active assignment using these thresholds with manual selection, we saw that fewer than 20% of the segments that were marked active manually were identified as active by these methods. We therefore iteratively lowered the threshold for each method until reaching a similar performance achieved by wavelet analysis (80-90% of the segments that were marked active manually were identified as active by the respective method). However, in doing so, we saw a sharp increase of false positive to ~20%.

**Electrophoresis.** λ-DNA, covalently labeled by rhodamine (Mirus) was embedded within 1.5 wt% agarose gel (Fisher, molecular biology grade, low EEO) in the presence of 1× TBE and glucose oxidase-based anti-photobleaching buffer. Imaging and tracking details are published elsewhere.[36] The strength of the electric field was 6 to 12 V/cm.

**Colloid glass.** Briefly, the system involves tracking modulated optical nanoprobes (MOON) tracer particles, prepared by coating a hemisphere of poly-methylmethacrylate (PMMA) particles with 12 nm of aluminum, in colloidal supercooled liquids comprised of PMMA colloids 1.42 μm in diameter at volume fraction 0.51. The solvent is index-matched and density-matched. Bright field imaging was used to track these probes as a function of time in four dimensions (x, y, in-plane and out-plane angles), the metal side facing the objective appearing black. Details of the experiments are published elsewhere.[39]

Acknowledgments. This work was supported by the U.S. Department of Energy, Division of Materials Science, under Award DEFG02-02ER46019.

**References:**

1. Kusumi, A.; Nakada, C.; Ritchie, K.; Murase, K.; Suzuki, K.; Murakoshi, H.; Kasai, R. S.; Kondo, J.; Fujiwara, T. Paradigm Shift of the Plasma Membrane Concept from the Two-Dimensional Continuum Fluid to the Partitioned Fluid: High-Speed Single-Molecule Tracking of Membrane Molecules. *Annu. Rev. Biophys. Biomol. Struct.* **2005**, *34*, 351-378.

2. Moerner, W. E. New Directions in Single-Molecule Imaging and Analysis. *Proc. Natl. Acad. Sci. U. S. A.* **2007**, *104*, 12596-12602.

3. Brandenburg, B.; Zhuang, X. Virus Trafficking - Learning from Single-Virus Tracking. *Nat. Rev. Microbiol.* **2007**, *5*, 197-208.

4. Waigh, T. A. Microrheology of Complex Fluids. *Rep. Prog. Phys.* **2005**, *68*, 685-742.

5. Wang, B.; Kuo, J.; Bae, S. C.; Granick, S. When Brownian Diffusion is not Gaussian. *Nat. Mater.* **2012**, *11*, 481-485.

6. Granick, S.; Bae, S. C.; Wang, B.; Kumar, S.; Guan, J.; Yu, C.; Chen, K.; Kuo, J. Single-Molecule Methods in Polymer Science. *J. Polym. Sci. Polym. Phys. Ed.* **2010**, *48*, 2542-2543.

7. Wong, I. Y.; Gardel, M. L.; Reichman, D. R.; Weeks, E. R.; Valentine, M. T.; Bausch, A. R.; Weitz, D. A. Anomalous Diffusion Probes Microstructure Dynamics of Entangled F-actin Networks. *Phys. Rev. Lett.* **2004**, *92*, 178101.

8. Cohen, A. E.; Moerner, W. E. Suppressing Brownian Motion of Individual Biomolecules in Solution. *Proc. Natl. Acad. Sci. U. S. A.* **2006**, *103*, 4362-4365.

9. Condamin, S.; Tejedor, V.; Voituriez, R.; Benichou, O.; Klafter, J. Probing Microscopic Origins of Confined Subdiffusion by First-Passage Observables. *Proc. Natl. Acad. Sci. U. S. A.* **2008**, *105*, 5675-5680.

10. Wang, B.; Guan, J.; Anthony, S. M.; Bae, S. C.; Schweizer, K. S.; Granick, S. Confining Potential when a Biopolymer Filament Reptates. *Phys. Rev. Lett.* **2010**, *104*, 118301.

11. Weeks, E. R.; Crocker, J. C.; Levitt, A. C.; Schofield, A.; Weitz, D. A. Three-Dimensional Direct Imaging of Structural Relaxation near the Colloidal Glass Transition. *Science* **2000**, *287*, 627-631.

12. Cheng, X.; McCoy, J. H.; Israelachvili, J. N.; Cohen, I. Imaging the Microscopic Structure of Shear Thinning and Thickening Colloidal Suspensions. *Science* **2011**, *333*, 1276-1279.

13. Turchin, P. *Quantitative Analysis of Movement: Measuring and Modeling Population Redistribution in Animals and Plants*. Sinauer Associates, Incorporated: 1998.

14. Zhao, K.; Tseng, B. S.; Beckerman, B.; Jin, F.; Gibiansky, M. L.; Harrison, J. J.; Luijten, E.; Parsek, M. R.; Wong, G. C. L. Psl Trails Guide Exploration and Microcolony Formation in Pseudomonas Aeruginosa Biofilms. *Nature* **2013**, *497*, 388-391.

15. Suh, J.; Wirtz, D.; Hanes, J. Efficient Active Transport of Gene Nanocarriers to the Cell Nucleus. *Proc. Natl. Acad. Sci. U. S. A.* **2003**, *100*, 3878-3882.

16. Jaqaman, K.; Loerke, D.; Mettlen, M.; Kuwata, H.; Grinstein, S.; Schmid, S. L.; Danuser, G. Robust Single-Particle Tracking in Live-Cell Time-Lapse Sequences. *Nat. Methods* **2008**, *5*, 695-702.

17. Zhu, L.; Zhang, W.; Elnatan, D.; Huang, B. Faster STORM Using Compressed Sensing. *Nat. Methods* **2012**, *9*, 721-723.

18. Sahl, S. J.; Leutenegger, M.; Hilbert, M.; Hell, S. W.; Eggeling, C. Fast Molecular Tracking Maps Nanoscale Dynamics of Plasma Membrane Lipids. *Proc. Natl. Acad. Sci. U. S. A.* **2010**, *107*, 6829-6834.

19. Liao, Y.; Yang, S. K.; Koh, K.; Matzger, A. J.; Biteen, J. S. Heterogeneous Single-Molecule Diffusion in One-, Two-, and Three-Dimensional Microporous Coordination Polymers: Directional, Trapped, and Immobile Guests. *Nano Lett.* **2012**, *12*, 3080-3085.

20. Arcizet, D.; Meier, B.; Sackmann, E.; Raedler, J. O.; Heinrich, D. Temporal Analysis of Active and Passive Transport in Living Cells. *Phys. Rev. Lett.* **2008**, *101,* 248103.

21. Monnier, N.; Guo, S.-M.; Mori, M.; He, J.; Lenart, P.; Bathe, M. Bayesian Approach to MSD-Based Analysis of Particle Motion in Live Cells. *Biophys. J.* **2012**, *103*, 616-626.

22. Bouzigues, C.; Dahan, M. Transient Directed Motions of GABA(A) Receptors in Growth Cones Detected by a Speed Correlation Index. *Biophys. J.* **2007**, *92*, 654-660.

23. Thompson, M. A.; Casolari, J. M.; Badieirostami, M.; Brown, P. O.; Moerner, W. E. Three-Dimensional Tracking of Single mRNA Particles in Saccharomyces cerevisiae Using a Double-Helix Point Spread Function. *Proc. Natl. Acad. Sci. U. S. A.* **2010**, *107*, 17864-17871.

24. Tuerkcan, S.; Alexandrou, A.; Masson, J.-B. A Bayesian Inference Scheme to Extract Diffusivity and Potential Fields from Confined Single-Molecule Trajectories. *Biophys. J.* **2012**, *102*, 2288-2298.

25. Huet, S.; Karatekin, E.; Tran, V. S.; Fanget, I.; Cribier, S.; Henry, J.-P. Analysis of Transient Behavior in Complex Trajectories: Application to Secretory Vesicle Dynamics. *Biophys. J.* **2006**, *91*, 3542-3559.

26. Zaliapin, I.; Semenova, I.; Kashina, A.; Rodionov, V. Multiscale Trend Analysis of Microtubule Transport in Melanophores. *Biophys. J.* **2005**, *88*, 4008-4016.

27. Tejedor, V.; Benichou, O.; Voituriez, R.; Jungmann, R.; Simmel, F.; Selhuber-Unkel, C.; Oddershede, L. B.; Metzler, R. Quantitative Analysis of Single Particle Trajectories: Mean Maximal Excursion Method. *Biophys. J.* **2010**, *98*, 1364-1372.

28. McKinney, S. A.; Joo, C.; Ha, T. Analysis of Single-Molecule FRET Trajectories Using Hidden Markov Modeling. *Biophys. J.* **2006**, *91*, 1941-1951.

29. Mallat, S. *A Wavelet Tour of Signal Processing*. Elsevier Science: 1999.

30. Percival, D. B.; Walden, A. T. *Wavelet Methods for Time Series Analysis*. Cambridge University Press: 2006.

31. Ivanov, P. C.; Rosenblum, M. G.; Peng, C. K.; Mietus, J.; Havlin, S.; Stanley, H. E.; Goldberger, A. L. Scaling Behaviour of Heartbeat Intervals Obtained by Wavelet-Based Time-Series Analysis. *Nature* **1996**, *383*, 323-327.

32. Yang, H. Detection and Characterization of Dynamical Heterogeneity in an Event Series Using Wavelet Correlation. *J. Chem. Phys.* **2008**, *129*, 074701.

33. Vale, R. D. The Molecular Motor Toolbox for Intracellular Transport. *Cell* **2003**, *112*, 467-480.

34. Mueller, M. J. I.; Klumpp, S.; Lipowsky, R. Tug-of-war as a Cooperative Mechanism for Bidirectional Cargo Transport by Molecular Motors. *Proc. Natl. Acad. Sci. U. S. A.* **2008**, *105*, 4609-4614.

35. Wang, B. Statistical Imaging of Transport in Complex Fluid: a Journey from Entangled Polymers to Living Cells. University of Illinois at Urbana-Champaign, 2011.

36. Guan, J.; Wang, B.; Bae, S. C.; Granick, S. Modular Stitching to Image Single-Molecule DNA Transport. *J. Am. Chem. Soc.* **2013**, *135*, 6006-6009.

37. Dorfman, K. D.; King, S. B.; Olson, D. W.; Thomas, J. D. P.; Tree, D. R. Beyond Gel Electrophoresis: Microfluidic Separations, Fluorescence Burst Analysis, and DNA Stretching. *Chem. Rev.* **2013**, *113*, 2584-2667.

38. Kegel, W. K.; van Blaaderen, A. Direct Observation of Dynamical Heterogeneities in Colloidal Hard-Sphere Suspensions. *Science* **2000**, *287*, 290-293.

39. Kim, M.; Anthony, S. M.; Bae, S. C.; Granick, S. Colloidal Rotation Near the Colloidal Glass Transition. *J. Chem. Phys.* **2011**, *135*, 054905.

40. Tokunaga, M.; Imamoto, N.; Sakata-Sogawa, K. Highly Inclined Thin Illumination Enables Clear Single-Molecule Imaging In Cells. *Nat. Methods* **2008**, *5*, 159-161.

41. Anthony, S.; Zhang, L. F.; Granick, S. Methods to Track Single-Molecule Trajectories. *Langmuir* **2006**, *22*, 5266-5272.

**Figures**

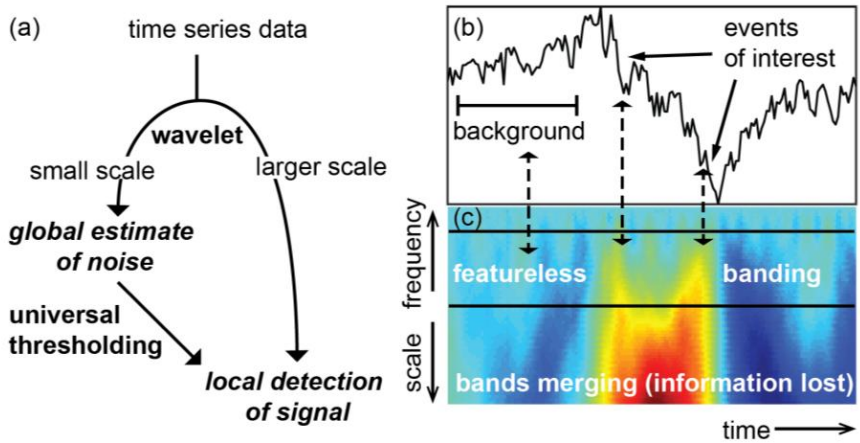

**Figure 1.** The main idea of wavelet analysis implemented in this study. (a) Schematically: the information embedded at small and large scales in a time series of raw data is expanded by wavelet transform. The information at small scale provides a global estimate of "noise" level and generates a universal threshold that can be projected onto long times, allowing one to localize "signal" from noisy background. (b) Spatial position plotted against time in an illustrative trajectory. (c) The corresponding wavelet coefficients are plotted against time as the result of local integration using the Haar continuous time wavelet transform (CWT). Red: positive values; blue: negative, green and yellow: near-zero. Dynamics of interest are identified on the scale (frequency) over which bands of coefficients are distinct. Information is lost if the scale is too large (frequency too small), whilst noise overwhelms events if the scale is too small (frequency too high). The usable range of scales is in between two black lines. Arrows in panel (b) point to two events of interest for further analysis. A time span of background noise is also highlighted in (b).

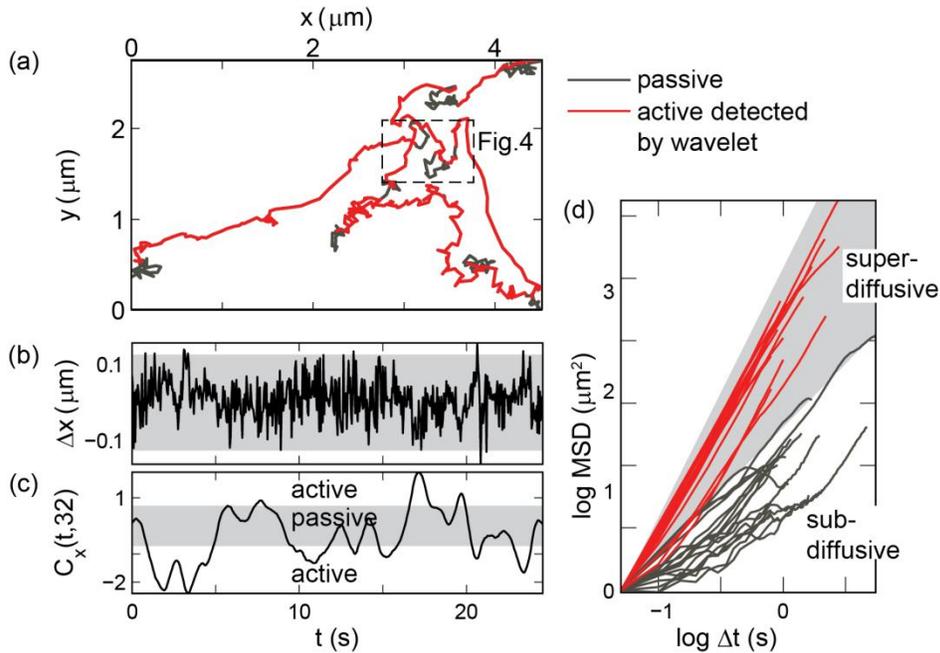

**Figure 2.** Example: a cell biology problem involving single-particle imaging of endosome transport along microtubules by molecular motors. (a) From a plot of the trajectory, x against y in Cartesian coordinates, wavelet analysis identifies active (red) and passive (gray) segments of the trajectory. (b) During this trajectory acquired by time-lapse imaging, the frame-to-frame displacement in the x direction (20 fps) is plotted against time. (c) The wavelet coefficients of this data at scale 32 frames are plotted against time, indicating the middle band of small coefficients that we identify with passive diffusion and the extreme values of wavelet coefficients that we identify with active motion. (d) Mean square displacement (MSD) is plotted against time on log-log scales for "active" (red) and "passive" (gray) segments of this trajectory. The shaded gray region demarcates the lower limit of Fickian diffusion with log-log slope 1, and upper limit of directional motion with log-log slope 2. These trajectories imputed from wavelet analysis split into two families, sub-diffusive (passive) and super-diffusive (active).

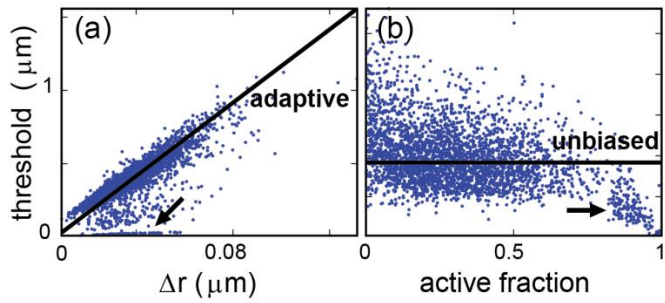

**Figure 3.** The universal threshold is adaptive and unbiased. (a) Continuing the cell biology example, for this scatter plot of 3443 endosome transport trajectories, each trajectory's threshold is determined individually and plotted against average frame-to-frame displacement of that trajectory. We conclude that universal thresholding is adaptive, as the threshold depends on the noise level of individual trajectory. (b) A scatter plot of the same data plotted against imputed active fraction of that trajectory. The horizontal black line, showing that the average threshold is uncorrelated with the imputed active fraction, suggests that the thresholds are unbiased by the active motion. Arrows note a subpopulation of trajectories (<10% of the total), for which universal thresholding needs refinement. The reasons for refinement are described in the text, and the text describes how to fine-tune the threshold iteratively and automatically.

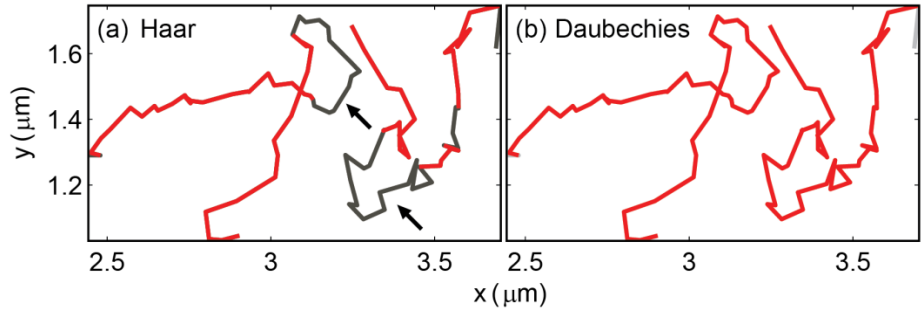

**Figure 4.** The choice of wavelet function depends on the physical problem. Continuing the cell biology example, a zone where the trajectory traces out loops corresponds to the boxed area in Fig. 2a. (a) Implementing the Haar wavelet, we assign a large portion of the loops (indicated by the arrows) as passive (gray), whereas (b) implementing the Daubechies 4 wavelet we assign them as active (red), but these two wavelets make the same identification elsewhere in the trajectory. The Haar wavelet is nonetheless preferred except for special circumstances owing to its simple, physical interpretation as a drift of mean position.

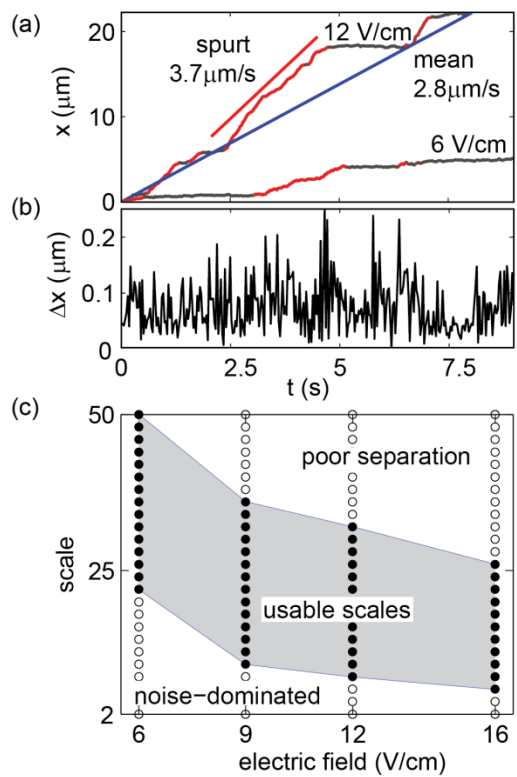

**Figure 5.** Example: a biotechnology problem involving single-molecule imaging of DNA electrophoresis in agarose gel as described in the text. (a) Illustrative trajectories showing that DNA center of mass motion at 12 V/cm is discontinuous, the wavelet analysis identifying spurts of rapid motion (red) and pauses between spurts (gray), neither of them equal to the mean speed. The trajectory at 6 V/cm drive is discontinuous likewise. (b) Frame-to-frame displacement of a 6V/cm trajectory (33 fps) is plotted against time. (c) This panel compares efficacy over a broad range of scale of analysis as well as drive voltage. Over a broad intermediate range of scale the mobility separation of this electrophoresis data is robust without depending on the specific choice of scale. Symbols represent examined conditions: solid, successful separation, open: poor separation.

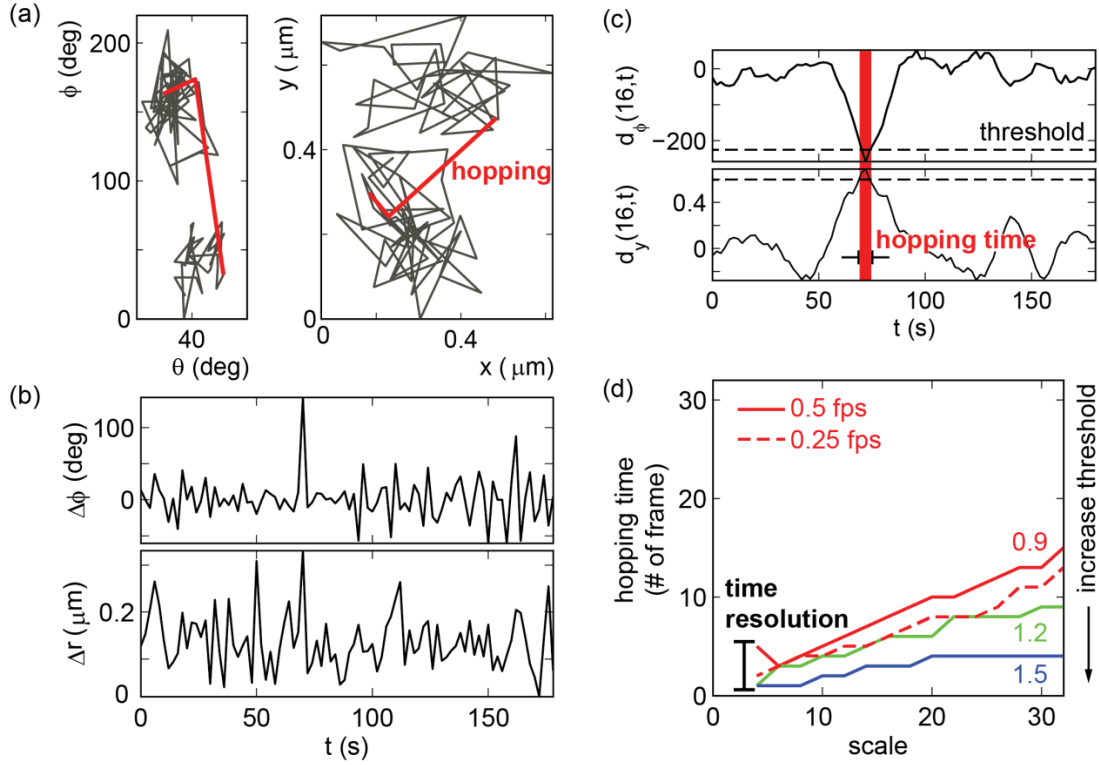

**Figure 6.** Example: a glassy dynamics problem involving the identification of hopping events. (a) Typical raw data showing an angular trajectory, $\theta$ plotted against $\phi$ (these are the out-of-plane and in-plane angles respectively) and the concomitant positional trajectory, x plotted against y, for the index-matched colloidal glass discussed in the text, showing the wavelet-identified hops (red) between regions of caged motion (gray). (b) Plotted against time, one observes the frame-to-frame in-plane angular and spatial displacement of these trajectories. (c) Coefficients of the wavelet transformation of the time series, evaluated at scale of 16 frames, are plotted against time for the in-plane angle and the y spatial position, each threshold denoted as a horizontal dashed line. The vertical red bar shows the interval when wavelet coefficients exceed the threshold, which identifies the hop. (d) Dependence of the imputed hopping time on the wavelet scale and threshold selected to analyze it. The threshold from universal thresholding is adjusted down or up by a constant factor of 0.9 (red), 1.2 (green), and 1.5 (blue).